\documentclass{aa}
\usepackage{graphicx}
\usepackage{txfonts}
\def\ha{\relax \ifmmode {\rm H}\alpha\else H$\alpha$\fi}

\def\nii{\relax \ifmmode {\rm N\,{\sc ii}}\else N\,{\sc ii}\fi}
\def\hii{\relax \ifmmode {\rm H\,{\sc ii}}\else H\,{\sc ii}\fi}
\def\hi{\relax \ifmmode {\rm H\,{\sc i}}\else H\,{\sc i}\fi}
\def\deg{\hbox{$^{\circ}$}}
\def\gtorder{\mathrel{\raise.3ex\hbox{$>$}\mkern-14mu
     \lower0.6ex\hbox{$\sim$}}}
\def\ltorder{\mathrel{\raise.3ex\hbox{$<$}\mkern-14mu
     \lower0.6ex\hbox{$\sim$}}}

\begin{document}

\title{Discovery of ultra-compact nuclear rings in three spiral galaxies
}

\author{S\'ebastien Comer\'on\inst{1} 
\and Johan~H.~Knapen\inst{1}
\and John~E.~Beckman\inst{1,2} 
\and Isaac~Shlosman\inst{3}
}

\offprints{S. Comer\'on}

\institute{Instituto de Astrof\'isica de Canarias, E-38200 La
Laguna, Spain
\and Consejo Superior de Investigaciones Cient\'ificas, Spain
\and Department of Physics and Astronomy, University of
Kentucky, Lexington, USA
}

\date{Received ; accepted}

\abstract{

Ring-shaped morphologies of nuclear star-forming regions within the
central $40-200$~pc of disk galaxies have been barely resolved so far
in three composite Sy~2 nuclei, the Sy~2 Circinus galaxy and in three
non-AGN galaxies. Such morphologies resemble those of the standard
$\sim 1$\,kpc-size nuclear rings that lie in the inner Lindblad
resonance regions of disk galaxies and, if they have a similar origin,
represent recent radial gas inflows tantalisingly close to the central
supermassive black holes. We aim to identify the population of such
ultra-compact nuclear rings (UCNRs) and study their properties in
relation to those of the host galaxies. From archival {\it Hubble
Space Telescope} UV and \ha\ images and from dust structure maps of
the circumnuclear regions in nearby galaxies, we analyse the morphology
of the star formation and dust, specifically searching for ring
structures on the smallest observable scales. In a sample of 38
galaxies studied, we have detected a total of four new UCNRs,
$30-130$~pc in radius, in three different galaxies.  Including our
confirmation of a previous UCNR detection, this yields a UCNR fraction
of roughly 10\%, although our sample is neither complete nor
unbiased. For the first time we resolve UCNRs in two LINERs. Overall
the UCNR phenomenon appears widespread and limited neither to late-type
galaxies nor exclusively to AGN hosts.


}
 
\maketitle

\section{Introduction} 

Star-forming nuclear rings, almost exclusively found in barred
galaxies (Buta \& Combes 1996), are attributed to shock focusing of
gas near the location of one or more inner Lindblad resonances (ILRs;
Schwarz 1984; Athanassoula 1992; Heller \& Shlosman 1996; see also Shlosman 1999;
Knapen 2005 for reviews).  Their sizes, of typically $\sim 1$\,kpc,
allow them to be easily resolved in ground- and space-based
observations. On the other hand, high resolution imaging with the {\it
Hubble Space Telescope (HST)} has led to the detection of three
unusually compact nuclear star-forming regions, where unambiguous
signatures of recent episodes of star formation (SF) coexist with the active galactic
nucleus (AGN)---so-called {\it composite} Sy~2 nuclei (Gonz{\'a}lez
Delgado et al. 1998). These barely resolved star-forming regions exhibit
morphologies which are not incompatible with those of ``standard''
nuclear rings---an issue that was never addressed in the
literature. The prevalence of such ``ultra-compact nuclear rings''
(UCNRs) and their origin are currently unknown. While conducting a
first comprehensive survey of UCNRs, supplemented by statistical,
phenomenological and dynamical studies of their origin and evolution,
we report here the results of the initial search of {\it HST} archival
imaging, which has revealed a number of UCNRs in NGC~2985, NGC~4579
and NGC~4800.
 
Besides the composite Sy~2 nuclei mentioned above, star-forming UCNRs
of less than 200\,pc in radius are known in the Sy~2 Circinus galaxy
(Wilson et al.~2000) and in galaxies without AGN: NGC~5248 (Laine et
al. 2001; Maoz et al. 2001; Jogee et al. 2002) and possibly in
NGC~3245 (Barth et al.~2001a) and IC~342 (B\"oker, F\"orster-Schreiber,
\& Genzel 1997).  The smallest of these UCNRs are around $\sim40$\,pc
in radius. Even more tantalising is the detection of molecular rings
or disks of $\sim 1-100$\,pc in low-luminosity AGN, such as M51 (Kohno
et al. 1996) or in NGC~1068 (Galliano et al. 2003). The Milky Way, a
``normal'' galaxy, hosts a massive molecular ring within $\sim 1.6 -
7$\,pc from the Galactic Centre.  Presently, it is not clear that all
these rings have a similar origin---but this is a possibility. A
major open question is whether the UCNR population is the small-radius
tail of the general population of nuclear rings (e.g., shown in Laine
et al. 2002) or whether it characterises a different physical
phenomenon.

Nuclear rings in general are taken to be by-products of gas inflow in
response to gravitational torques induced by bars or galaxy
interactions, so it is significant that rings are found so
close to the region dominated by the supermassive black hole (SMBH). 
UCNR hosts are nuclear starburst galaxies (by definition)
and potentially related to the triggering and fuelling of AGN.

There are two closely related scenarios that could account for the tiny 
UCNRs compared to their ``normal''
counterparts, namely, a shallower radial mass distribution in the
central kpc of the host galaxy, implying a smaller bulge, or 
ring  formation in response to a non-axisymmetric mass distribution in 
the inner few hundred pc, that tumbles exceedingly fast. In both cases this 
implies that the ILR region is located close to the centre.  
The first scenario would indicate that one might expect a
deficiency of UCNRs in early-type spirals. The second scenario
favours the presence of nuclear gaseous or stellar bars (Shlosman 1999). 
These alternatives can be distinguished using a number of observational
tests.

\section{Observations and data reduction}

We inspected {\it HST} archival images of a sample of 38
nearby spiral galaxies, all at a distance of less than 50 Mpc, for
which images in \ha\ and/or in UV bands (filters F218W, F250W, F300W,
or F330W) were available. The sample was selected to
study the possible relationships between central drops in the stellar
velocity dispersion (``$\sigma$-drops'') and the host morphology on the
smallest observable scales, and consists of 19 galaxies with confirmed
$\sigma$-drops as known from the literature and 19 galaxies in a
control sample, matched carefully to the $\sigma$-drop sample in terms
of absolute magnitude, distance, axial ratio and morphological type
(S. Comer\'on et al. 2007, in preparation). Highly inclined 
galaxies were
excluded. We used images from both ACS, with 0.05\,arcsec resolution,
and WFPC2, with 0.0455\,arcsec resolution.  Cosmic rays were
removed with the {\sc qzap} algorithm.

We focus here on the three galaxies in which we discovered UCNRs. For
NGC~2985, we combined two UV 1400\,s WFPC2 F218W exposures of a field
centred on the nucleus of the galaxy. An \ha{} image was obtained
through the F658N filter with ACS, while we used an ACS F814W image
for continuum subtraction. We used the ACS F814W image and a WFPC2
F606W image to construct a colour index image. For NGC~4579 we
combined two WFPC2 F791W 300\,s exposures to subtract the continuum
from an ACS set of three combined F658N images. We also used two sets
of two UV ACS 300\,s exposures, taken through the F250W and F330W
filters. For the colour index image we used the WFPC2 F791W images and
a set of two F547M WFPC2 images that we combined. For NGC~4800, we used
a 1000\,s F300W WFPC2 exposure, a pair of ACS F658N and F814W images
for \ha{}, and F606W and F814W WFPC2 images in order to build a colour
index image. We used standard image reduction software, and followed
the recipe described by Knapen et al.  (2006) for the H$\alpha$
continuum subtraction. The resulting UV and H$\alpha$ images of the
three galaxies are shown in Fig.~1 ({\it top} and {\it centre-top}
panels, respectively). Colour index images are in the {\it bottom}
panel.

We produced \textquotedblleft structure maps" (Pogge \& Martini 2002) 
from the broad-band red images to bring out the distribution of dust 
in the central regions of the sample galaxies, which are enhanced on 
the scale of the point spread function (PSF). We used synthetic PSFs, created with the
\textquotedblleft Tiny Tim" software (Krist \& Hook 1999) The structure 
maps are shown in the {\it centre-bottom} panels of Figure~1.

\section{Results}

\subsection{NGC 2985}

NGC~2985 is an SAab galaxy at $D=22.4$\,Mpc (Tully 1988).
In the UV and \ha{} images (Fig.~1) we can readily see a ring of emission,
caused by the presence of massive SF. The ring is well defined:
the intensity falls off sharply both within and outside
the ring. There is no equivalent in the F814W band which is a further
indication that the ring is due to a young stellar component. In both
UV and \ha\ the radius of the ring is approximately 0.5\,arcsec or 
50\,pc. The intense peak of emission at the nucleus is due to a LINER 
(Wilner et al.~1985). It is more extended in \ha{}, with a half 
intensity radius of close to 25\,pc,
than in the UV band. The UCNR is not uniform azimuthally, but quite
patchy, being clearly composed of a number of individual massive star
clusters. Each cluster emits in UV and \ha, though the structure
in UV is finer, as would be expected since the \ha{} comes from the
more extended ionised gas around the clusters. The structure map shows
dust lanes spiralling in towards the central region of the
galaxy, but fading out as they approach the radius of the UCNR. These
dust lanes are not necessarily related to material flowing towards the 
centre and can be formed from a mild compression not accompanied by SF 
(Laine et al. 1999; Englmaier \& Shlosman 2000). Some
dust structure inside the ring may hint at the presence of a
residual flow towards the nucleus, which is compatible with the
nuclear SF peak.

This galaxy does not show a sign of a bar even on the
smallest scales ({\it HST} near-IR imaging: Laine et
al. 2002) even though  it is possible  that deeper imaging will reveal one. Gravitational torques triggered by a minor merger are an
alternative way to induce inflow (see Knapen et al. 2004, 2006 for
examples). Indeed NGC~2985 ($B=11.2$\,mag) has two close
neighbours. NGC~3027 is a 12.2\,mag SB(rs)d galaxy at projected
distance 160\,kpc, with a systemic velocity difference of only
300\,km\,s$^{-1}$. The 17.2\,mag irregular galaxy KDG~59
(Karatchentseva et al. 1987) is closer at 90\,kpc and $\Delta
v=140$\,km\,s$^{-1}$, but its faintness implies that its gravitational
influence on NGC~2985 is $\sim 10$ times less than that of
NGC~3027. H{\sc i} observations clearly show a morphological and
kinematical disturbance of NGC~2985, possibly due to KDG~59, but there
is no evidence for a tidal tail which connects to NGC~3027
(Noordermeer et al. 2005; T. A. Oosterloo, private communication). We conclude that
one or both of these companions can cause a deviation from axisymmetry
in the centre of NGC~2985, which in turn can trigger the observed
UCNRs.

\subsection{NGC 4579}

NGC~4579  has been studied extensively. It has been classified
as SABb in the RC3, and as SBab and SBa in the $B$ and $H$-bands by
Eskridge et al. (2002). Garc\'\i a-Burillo et al. (2005) 
reported a 1.5\,arcsec diameter
ring-like structure in $V-I$ images of NGC~4579, which corresponds to
a radius of 80\,pc at $D=16.8$\,Mpc (Tully 1988). On the
basis of our UV image (Fig.~1), we can say that it is not a true ring. Our UCNR is not centred on the
AGN nucleus (Seyfert 1.9: Maiolino et al. 1997, or LINER: Barth et al. 2001b).
Upon close inspection, the \ha\ image reveals a very diffuse
ring shape following the UCNR which is defined more clearly in the UV.
A set of much more brightly emitting arc-shaped regions is seen, well
within the UCNR, partly traced by dust as outlined in the structure
map in Fig.~1, and possibly related to the outflow from the AGN
(see Section 4).
The inner arcs have a semi-major axis of
about 80\,pc and a position angle (PA) of 69\deg, the ring's radius is 135\,pc and
PA=89\deg. The latter is approximately the same as that of the galaxy
(PA=95\deg).  The southern sections of the ring and the arcs are
almost coincident (Fig.~1). Part of the inner arcs coincides with the
structure that was suggested by Garc\'\i a-Burillo et al. (2005), 
but they gave a very different PA.

The UCNR so clearly visible in the UV image of
NGC~4579 is clearly off-centred, which can be readily explained in
terms of a resonance ring in a mass distribution that is induced by 
a superposition of $m=1$ and $m=2$ perturbations. 

The inner arcs seen in the UV can originate in the AGN
activity detected in this low-luminosity LINER (or Sy) galaxy.
The line emission in the central 100\,pc can be reconciled with the
presence of a jet and emitted by
$\sim100\,{\rm km\,s^{-1}}$ shocks triggered by the interaction of the
compact radio jet with the cloudy ambient gas (Contini 2004).
SF can be induced in dense clumps embedded in the
jet backflow (cocoon). Radio surveys of
low-luminosity AGN find that low radiative efficiency accretion flows
in these objects can coexist with the collimated outflows (e.g.,
Doi et al. 2005). In particular, the PA of the jet in NGC~4579
(Sofue et al. 2004; Doi et al. 2005), 
$\sim57^{\rm o}-65^{\rm o}$, is well aligned with the PA of the inner
arcs, $\sim69^{\rm o}\pm 20^{\rm o}$, shown in Fig.~1, which probably is the
orientation of the ionisation cone in this AGN. Garc\'\i
a-Burillo et al. (2005) find signatures of outflow velocities in this
region, thus indirectly supporting this picture. While such
low-luminosity AGN can lack fully developed molecular
tori, their radio loudness is known to increase with a decreasing
accretion rate (Ho 2002; Greene, Ho, \& Ulvestad 2006). This effect is apparently
related to AGN switching their outflows from molecular tori at high
luminosities to jets at lower ones (Elitzur \& Shlosman 2006).

\subsection{NGC 4800}

NGC~4800 is an SAb galaxy at $D=15.2$\,Mpc (Tully 1988).  A
double-ring structure in the circumnuclear zone is seen in the UV
and \ha\ images (Fig.~1). The inner of these two rings, with a
radius of only about 30\,pc, is very close to the bright nucleus
(classified as \hii) and is rather faint and patchy. The outer
ring, with a radius of $\sim 130$\,pc, is much brighter and
wider. A dust trail covers a part of its perimeter. The structure map
shows dust spiralling inwards until the outer edge of the bigger ring.
The lack of structure between the rings implies a lack of ordered dust.

Profile analysis of a 2MASS near-IR image of
NGC~4800 indicates the presence of a weak bar (with a deprojected
ellipticity some 0.2 higher than the disk) with a semi-major axis of
some 35\,arcsec, although it is classified as a non-barred galaxy. The
fact that NGC~4800 has an exponential ``bulge'' (Andredakis \& Sanders
1994) argues in favour of bar-induced dynamics in the central region
of this galaxy. NGC~4800 does not
have a close companion, and the \hi\ morphology, which could yield
information on a possible interacting past (e.g., Knapen et al. 2004) is
unknown. 

Photometric analysis of the UCNRs indicate that they are likely to be highly obscured by dust. This obscuration is probably greater than two magnitudes in \ha{} and may be specially important in the case of the UCNRs of NGC~4800.

\section{Star formation rates}

We have estimated star formation rates (SFRs) for the UCNRs from the
H$\alpha$ and UV, using the empirical formulae given by Kennicutt
(1998). We could not apply these in the case of the UV image of
NGC~4800 because its wavelength is outside of the range of the
Kennicutt formula, which is 1500$-$2800\,\AA. We assumed a
conservative dust extinction value of
$A_{\rm{H}\alpha}=A_{\rm{UV}}=2\,\rm{mag}$, and find values for the
SFRs which are around
$0.04\,M_{\odot}\,\rm{year}^{-1}$ (Table~1). We find that
the SFRs as derived from the H$\alpha$ and UV images are of the same order of magnitude for those galaxies
where we could derive the UV SFR. This is surprising given that we
used the same extinction correction in both cases, but can be
explained by one or, most plausibly, a combination of the following arguments.

Firstly, and as a most plausible explanation, at the rather modest
SFRs we find in the UCNRs, it is likely that the upper end of the
Salpeter IMF assumed by Kennicutt (1998) is not fully populated.
Purely due to low-number statistics, there are too few O stars to
produce H$\alpha$ emission commensurate with the SFR. In contrast, the
low-mass (e.g., B-type) stars which produce UV emission are much more
common and will be present in sufficient numbers. This means that the
H$\alpha$ SFR may be underestimated by an unknown amount. Secondly,
the UCNRs may be coeval and rather old, so that the H$\alpha$ emission
due to the most massive stars is fading, but the UV not yet.

This leads to the conclusion that the true SFR is
underestimated by the numbers in Table~1. For instance, in the first
option, we would use only the UV SFR, and in that case an $A_V$ of
2\,mag seems very conservative indeed, given that the UCNRs are so
close to the nucleus. Assuming an $A_V$ of 3\,mag already yields a SFR
close to $0.1\,M_{\odot}\,\rm{year}^{-1}$. It is, unfortunately, not
feasible to reach reliable determinations of $A_V$ from, e.g., the
colour index images in Fig.~1, because of the combined reddening
effect of extra dust and a relatively older stellar population. We
thus conclude that a reasonable estimate of the SFR of a typical UCNR
is $0.1\,M_{\odot}\,\rm{year}^{-1}$. Given the small size of the
UCNRs, the SFR densities, or SFR per unit area (see Table~1) are, at a
value of order $10^{-6}\,M_{\odot}\,\rm{yr}^{-1}\,\rm{pc}^{-2}$,
comparable to that in the starburst region in M83 (Harris et
al. 2001).  If UCNRs are even moderately long-lived and stable, this
is a significant SFR, which may transform important quantities of gas
into stars, very close to the nuclei of their host galaxies. For
instance, this SFR during $10^8$\,yr would yield a mass transformed
from gas into stars of $10^7\,M_{\odot}$, which would add to the inner
bulge mass and thus assist in secular evolution.

\begin{table*}
\begin{center}
\begin{tabular}{l c c c c}
Galaxy & $\rm{H}\alpha$  $(M_{\odot}\ \rm{yr}^{-1})$ &  UV  $(M_{\odot}\ \rm{yr}^{-1})$ & $\rm{H}\alpha$  $(M_{\odot}\ \rm{yr}^{-1} \rm{pc}^{-2})$ & UV $(M_{\odot}\ \rm{yr}^{-1} \rm{pc}^{-2})$\\
\hline
NGC~2985 & 0.026 & 0.050 & $1.8 \times 10^{-6}$ & $3.4 \times 10^{-6}$ \\
NGC~4579 & 0.042 & 0.064 & $1.5 \times 10^{-6}$ & $2.3 \times 10^{-6}$ \\
NGC~4800 (inner UCNR) & 0.00063 & $-$  & $2.7 \times 10^{-7}$ & $-$\\
NGC~4800 (outer UCNR) & 0.030 & $-$  & $6.0 \times 10^{-7}$ & $-$ \\
\hline
\end{tabular}
\caption{\label{table} SFRs (columns 2 and 3) and SFR densities
  (columns 4 and 5) of the discovered UCNR, as determined from
  the H$\alpha$ and UV images. Foreground dust extinction of
  $A_{{\rm H}\alpha}=A_{\rm UV}=2$\,mag was assumed in all cases.}
\end{center}
\end{table*}

\begin{figure*}
\begin{center}
\begin{tabular}{c}
\includegraphics[width=0.25\textwidth]{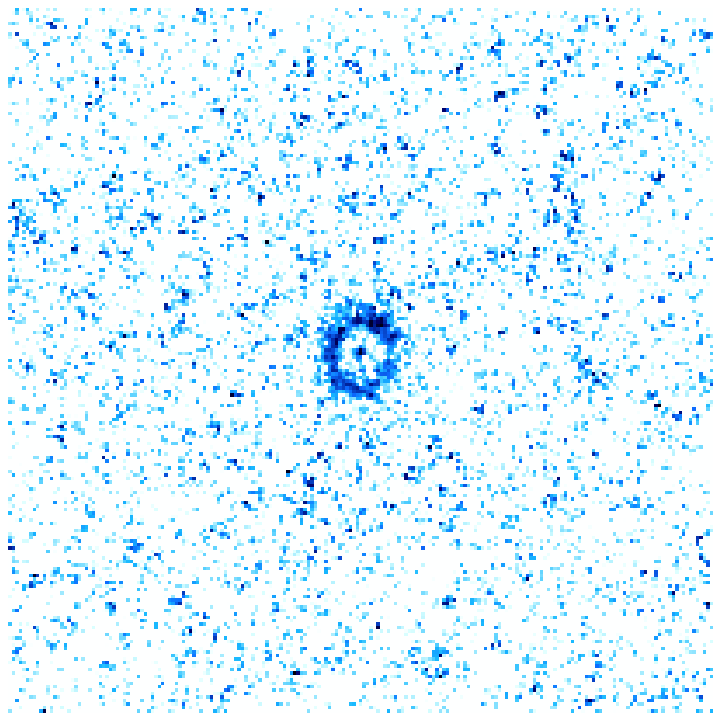}
\includegraphics[width=0.25\textwidth]{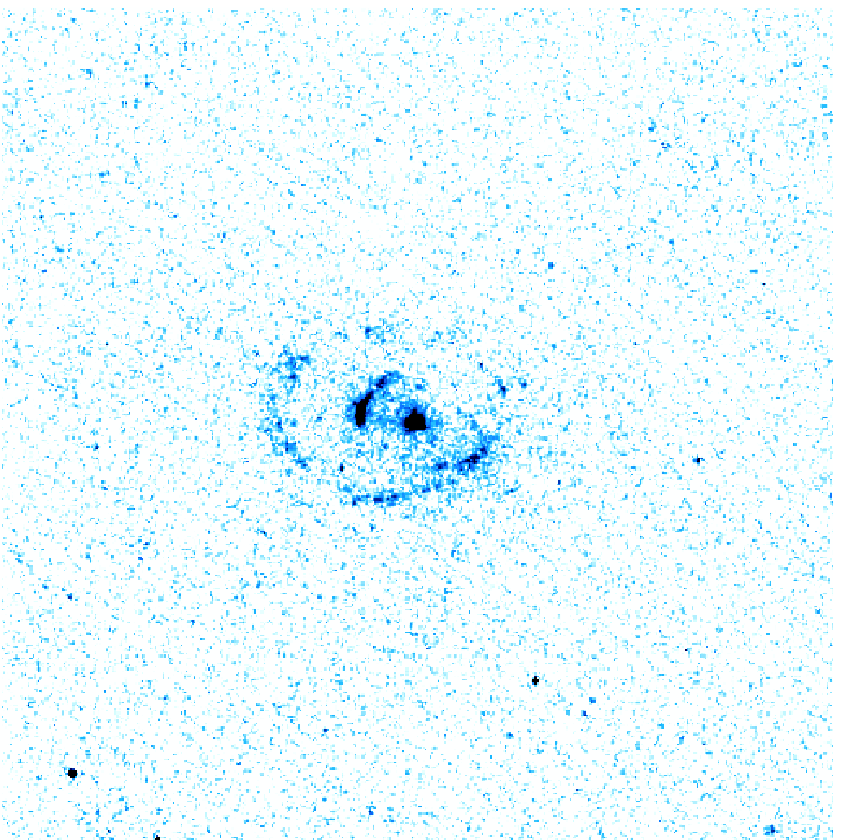}
\includegraphics[width=0.25\textwidth]{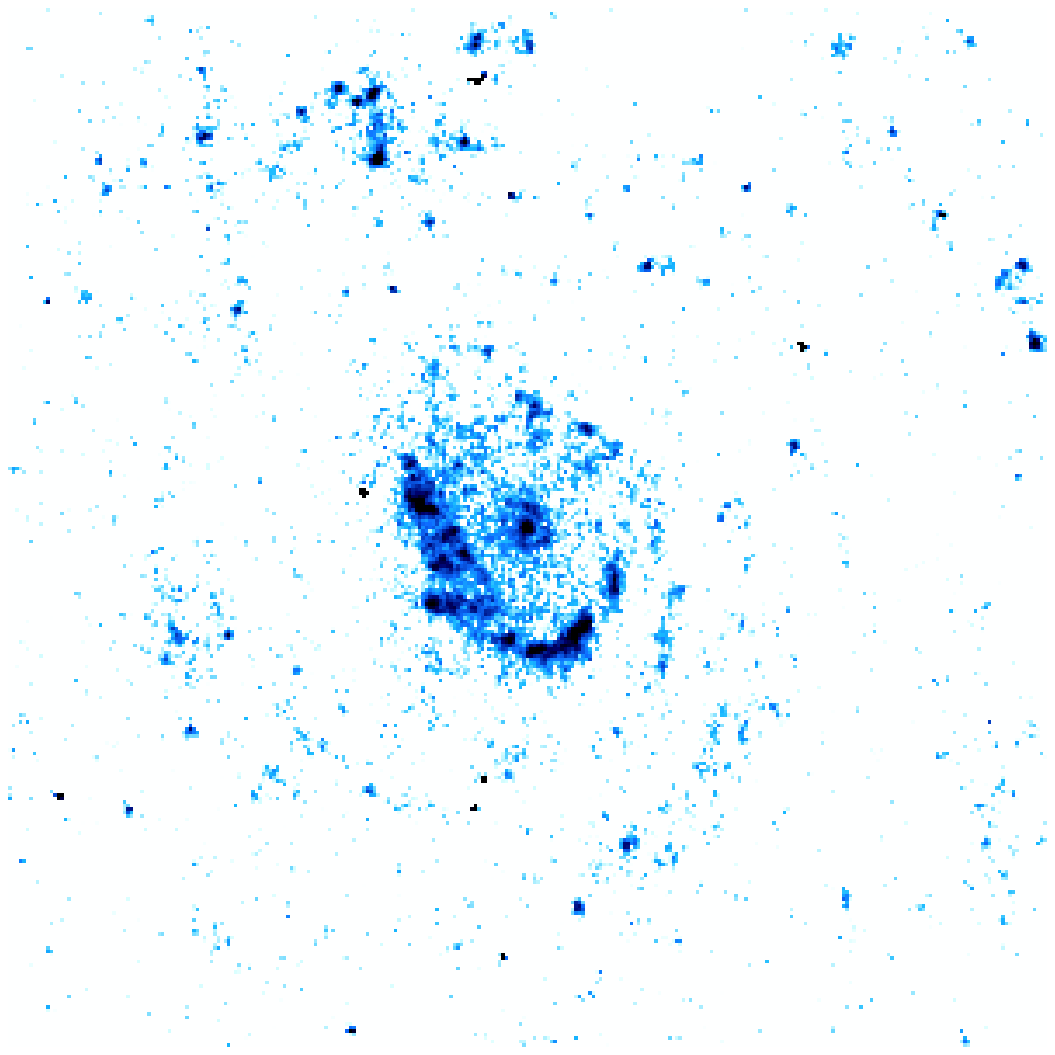}\\
\includegraphics[width=0.25\textwidth]{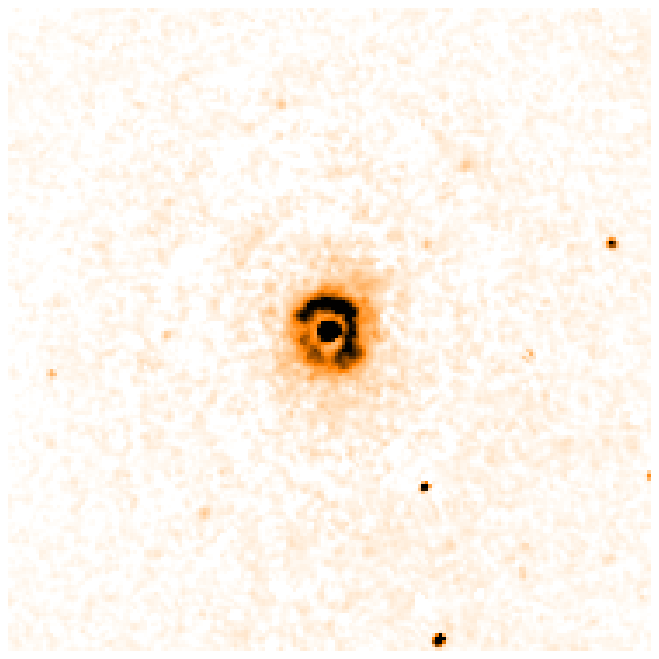}
\includegraphics[width=0.25\textwidth]{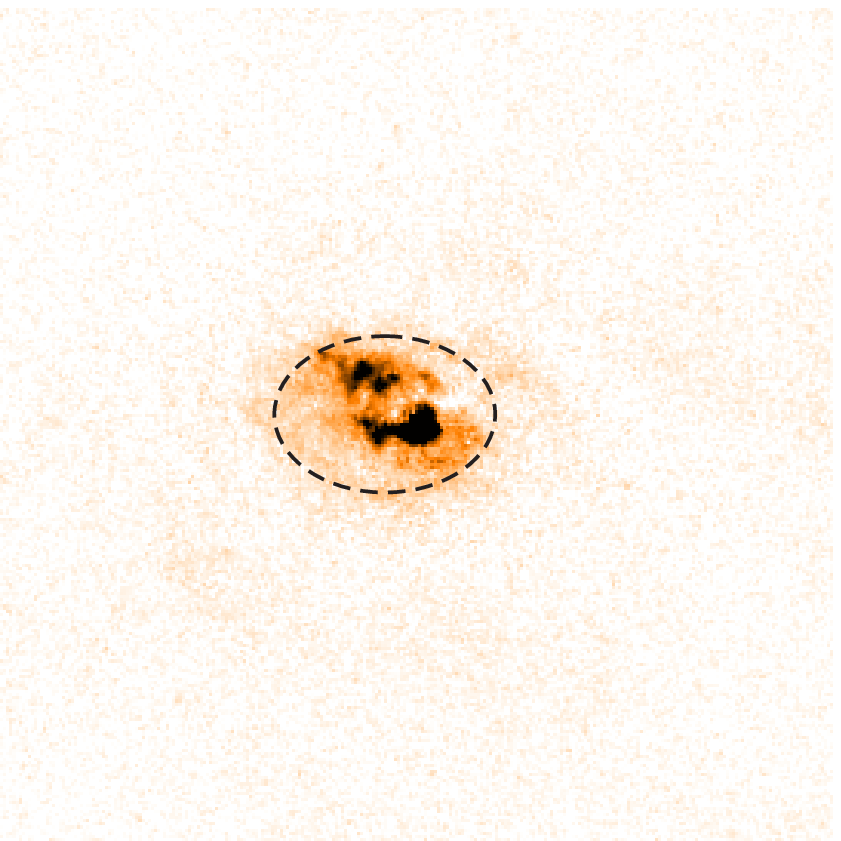}
\includegraphics[width=0.25\textwidth]{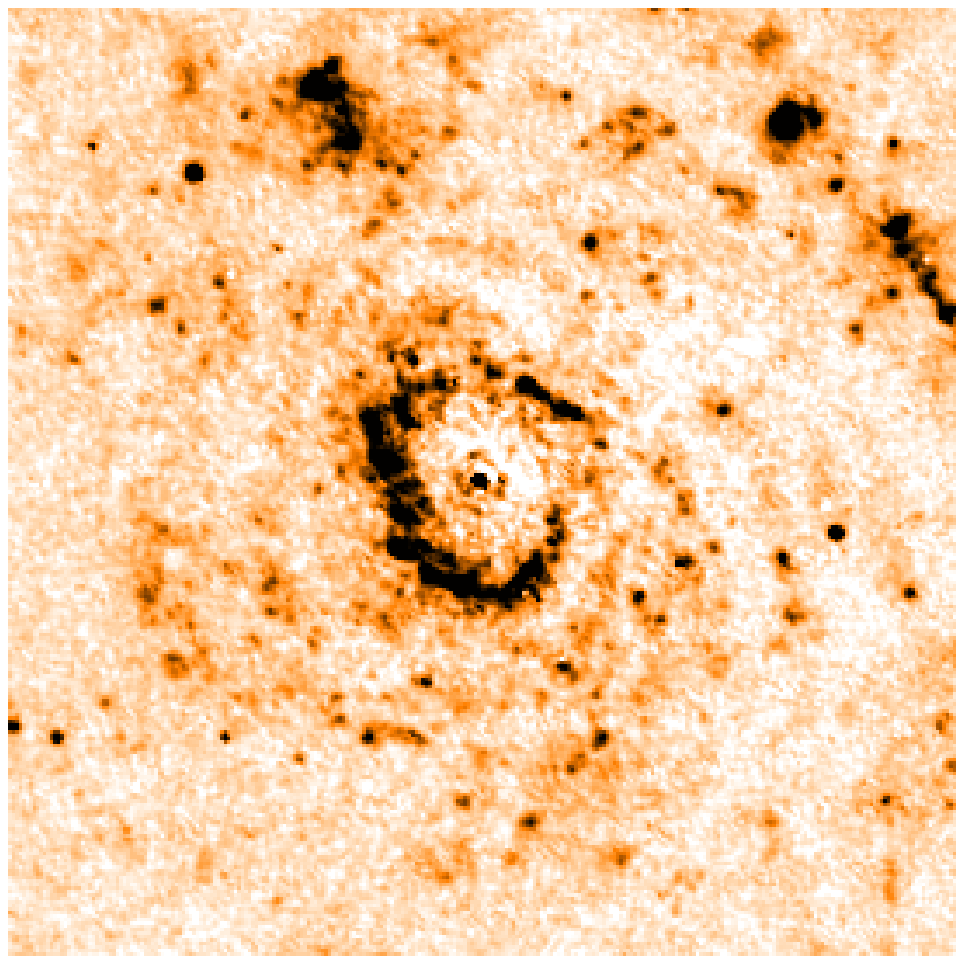}\\
\includegraphics[width=0.25\textwidth]{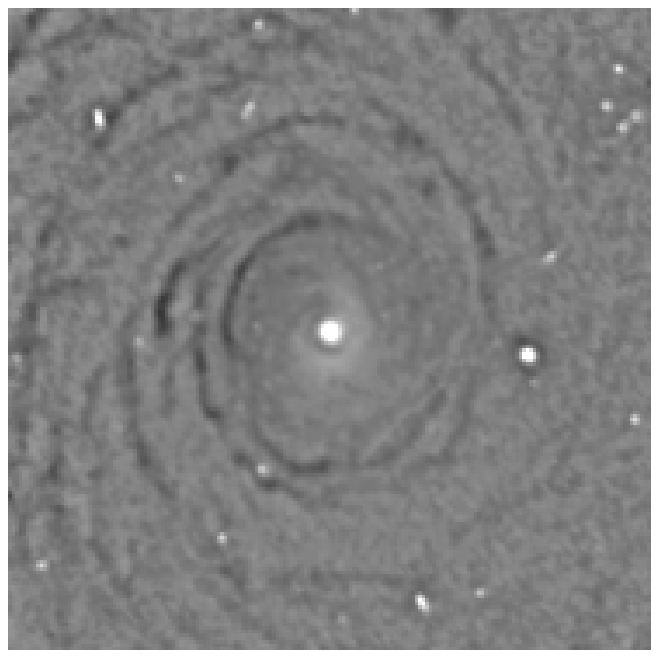}
\includegraphics[width=0.25\textwidth]{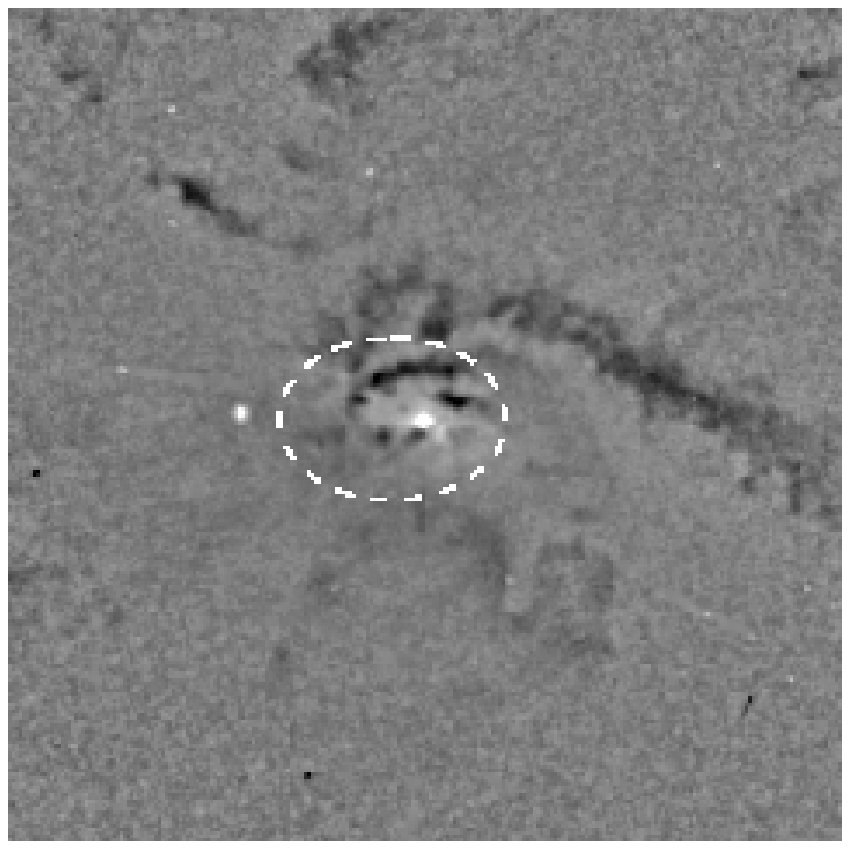}
\includegraphics[width=0.25\textwidth]{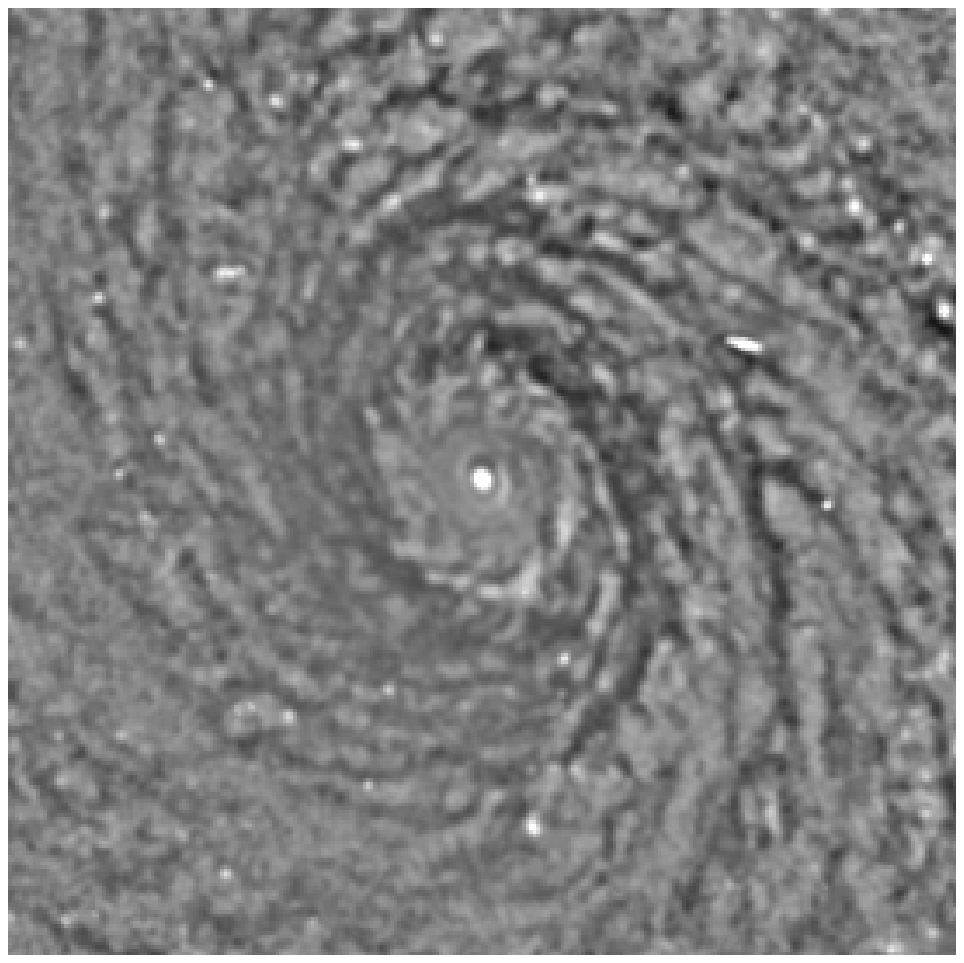}\\
\includegraphics[width=0.25\textwidth]{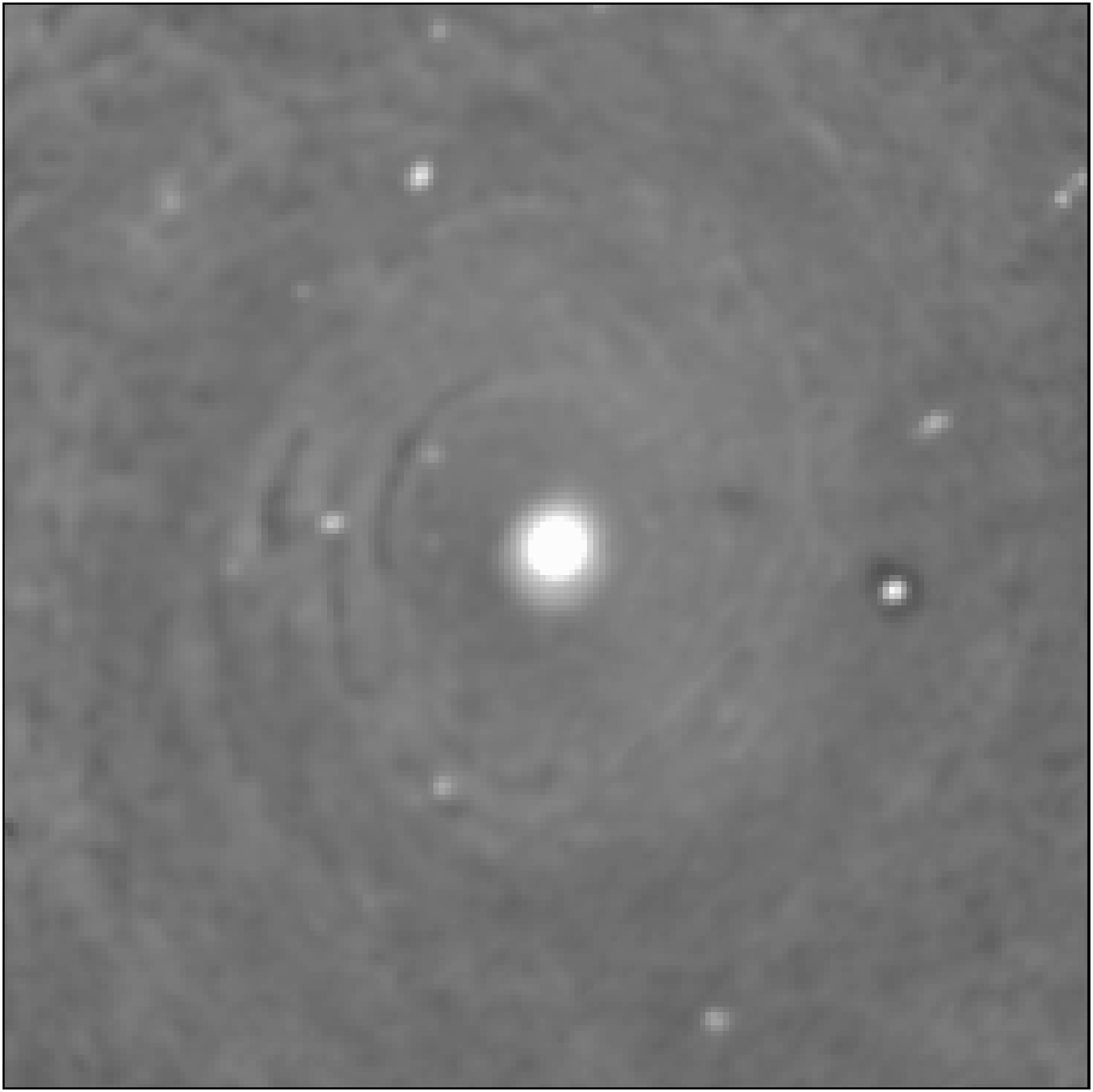}
\includegraphics[width=0.25\textwidth]{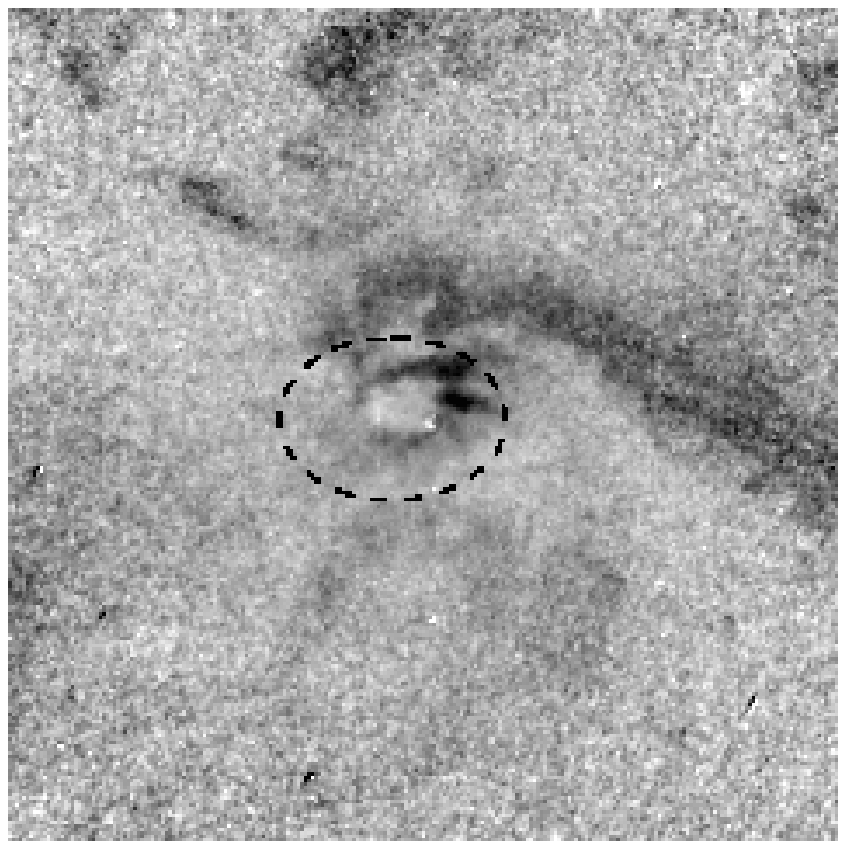}
\includegraphics[width=0.25\textwidth]{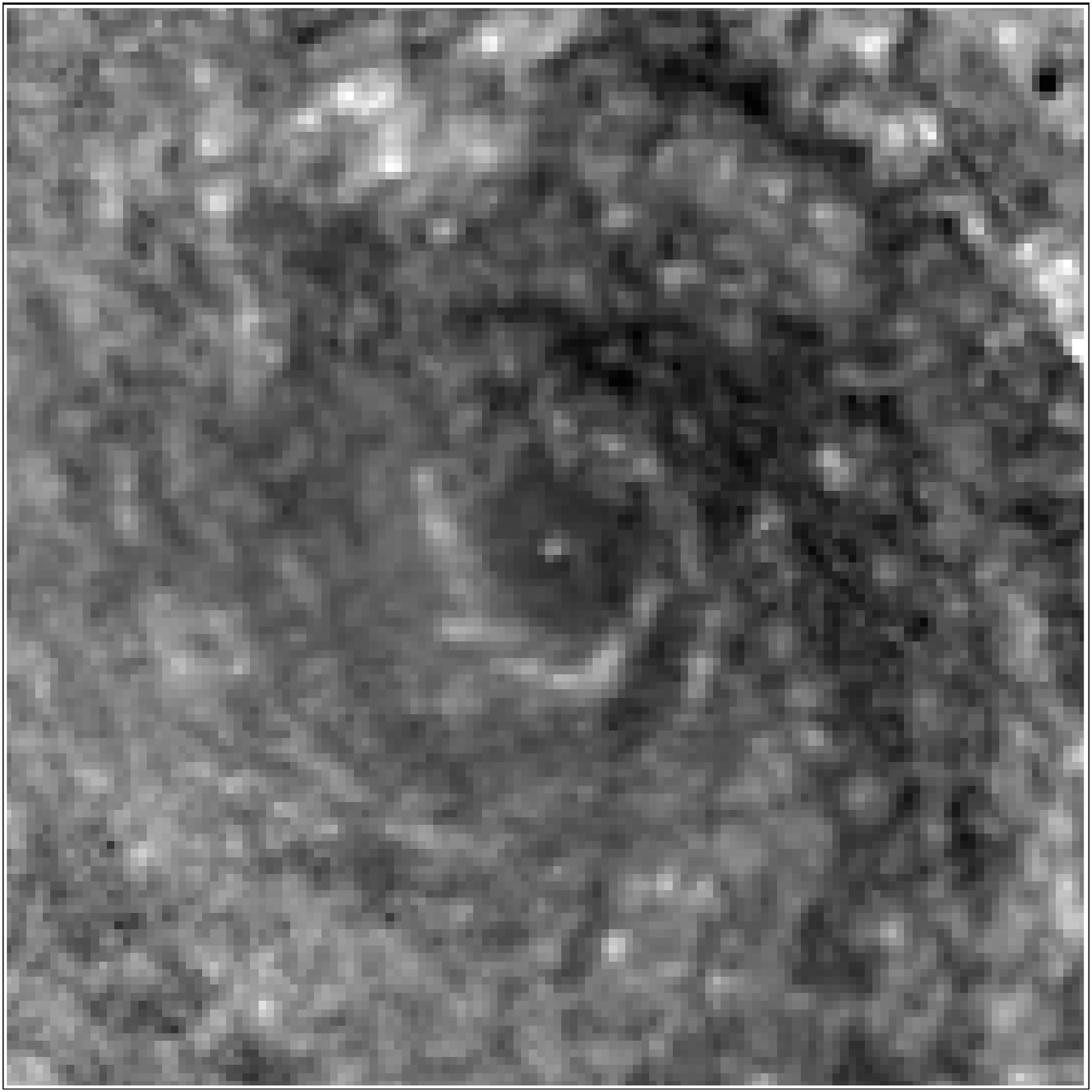}\\
\end{tabular}

\caption{\label{images} Images of the central regions of the three
galaxies of our study (NGC~2985, {\it left}; NGC~4579, {\it centre};
and NGC~4800, {\it right}). Images in the {\it top} row are in the UV
(filters F218W, F250W and F300W, from left to right), the {\it
central-top} row is \ha, the {\it central-bottom} row shows the structure
maps and the {\it bottom} row show colour index images (F606W-F814W,
F547M-F791W and F606W-F791W, from left to right). All images span
1\,kpc on the side and the scales are such that this corresponds to
9\,arcsec for NGC~2985, 12\,arcsec for NGC~4579, and 13.5\,arcsec for
NGC~4800. The shape of the UCNR in NGC~4579, as derived from the UV image, is outlined with ellipses. North is up and East to the left.}
\end{center}
\end{figure*}

\begin{figure*}
\begin{center}
\begin{tabular}{c}
\includegraphics[width=0.25\textwidth]{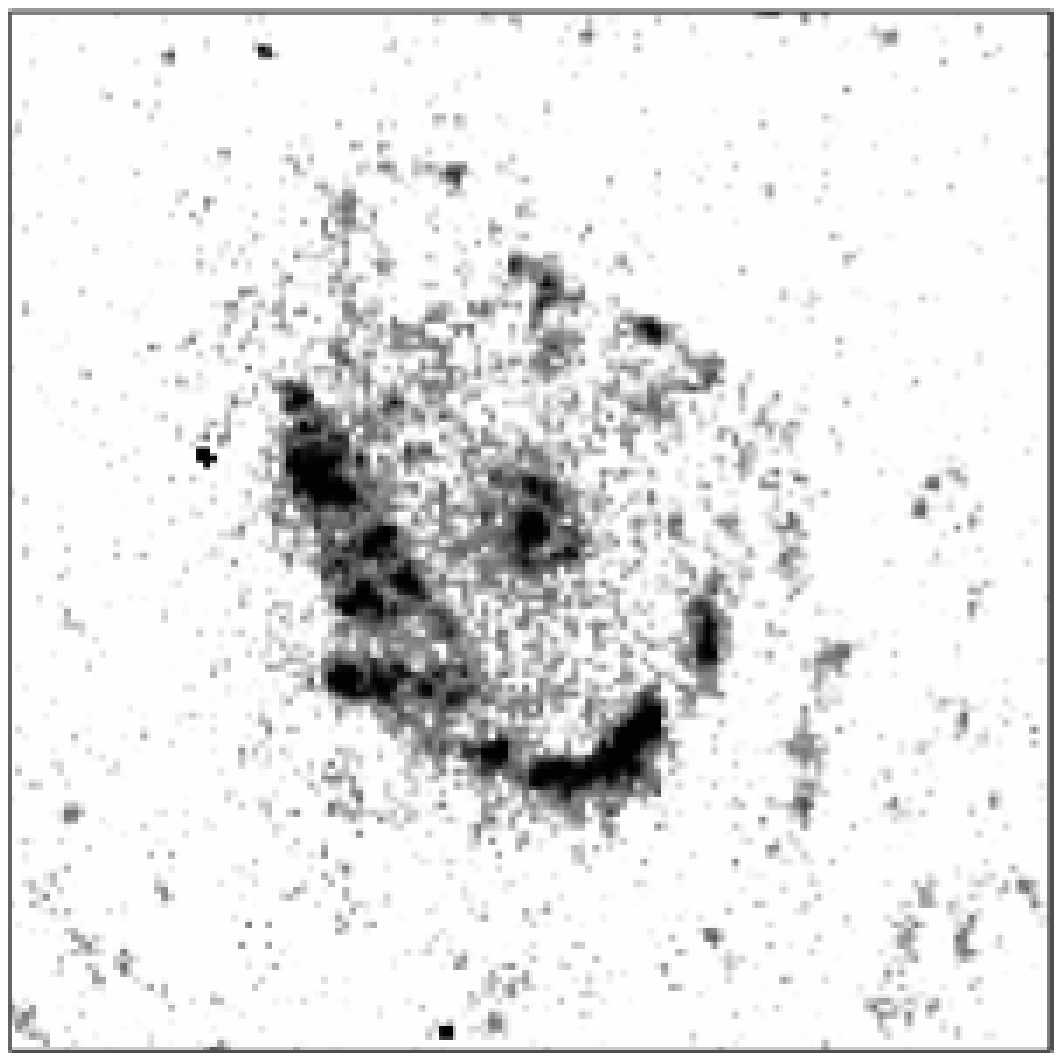}
\includegraphics[width=0.25\textwidth]{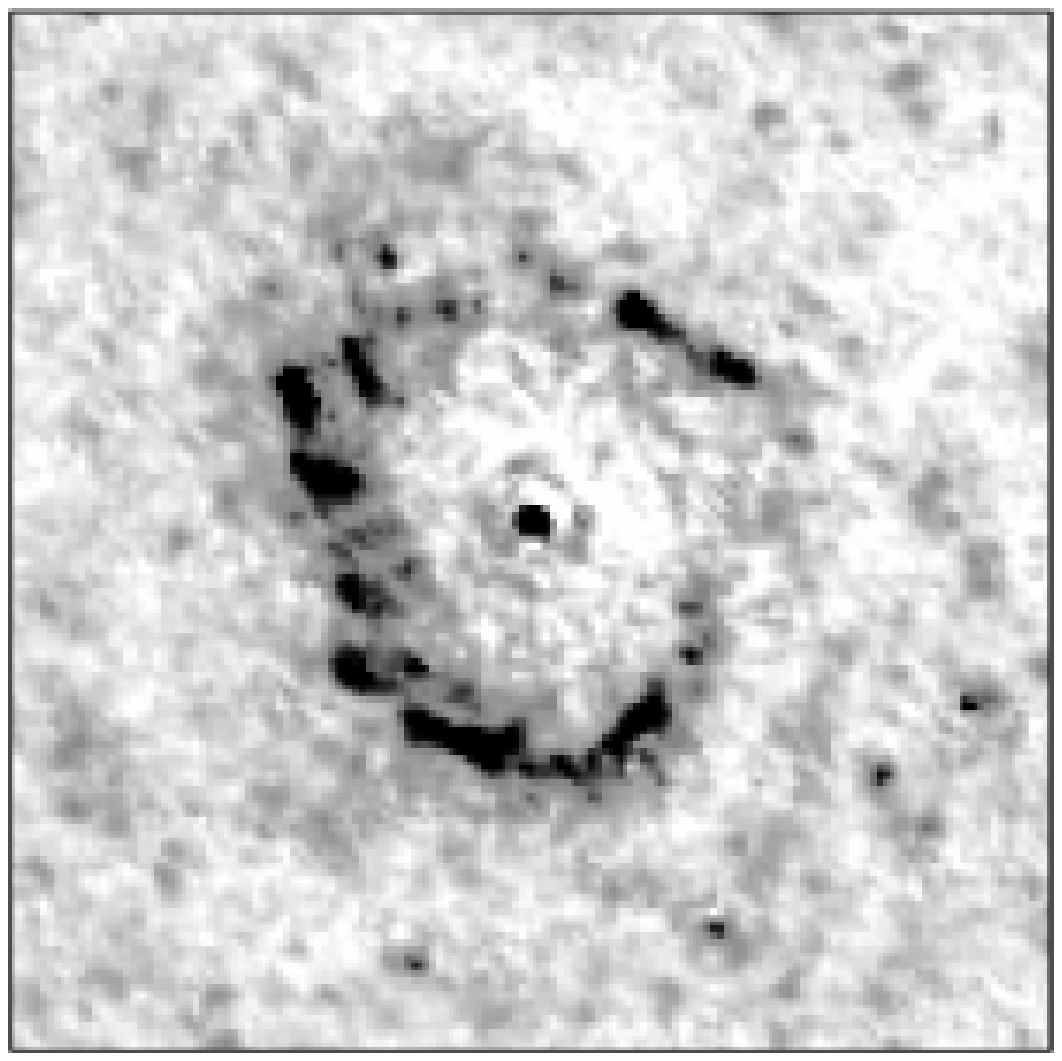}
\includegraphics[width=0.25\textwidth]{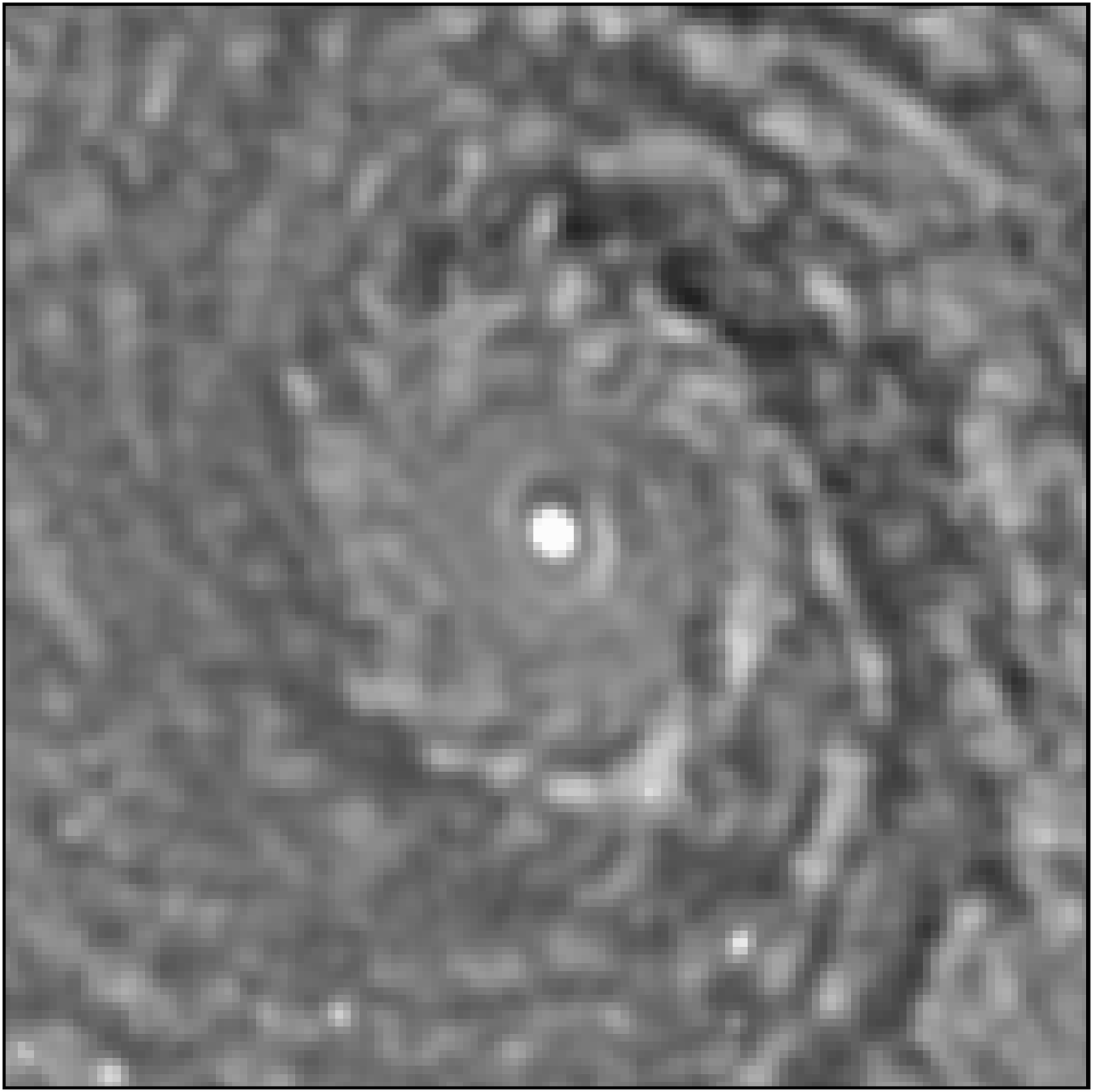}
\end{tabular}
\caption{\label{zoom} Close-up of the images of NGC~4800
  (Fig.~\ref{images}). From {\it left} to {\it right} UV, \ha{}, and the structure map. The colour index image has much poorer spatial resolution and is thus not included here. All images span 0.5\,kpc on the side, which corresponds to 6.7\,arcsec. North is up and East to the left.}
\end{center}
\end{figure*}
 
\section{Discussion}

Careful analysis of a large, statistically significant sample is necessary to 
quantify the significance of UCNRs and the relations to their host
galaxies (in preparation). Here we report the
detection of four new UCNRs in a non-AGN (NGC~4800) and in two AGN hosts
(NGC~2985 and NGC~4579). The latter provide the first detection of UCNRs in 
LINERs. The number of known UCNRs, nearly doubled with the present
study, and their appearance 
among early and late Hubble types, as well as in AGN and non-AGN hosts, 
points to a common phenomenon in galaxy evolution. 

The observed morphologies of the star-forming regions
in three Sy~2 composite nuclei (see Section~1) are compatible with those
of standard kpc-size nuclear rings. Because the 
composite nuclei are estimated to comprise about 50\% of the Sy~2 nuclei, 
this can provide a lower limit for their spread.

In addition to the new detections, our sample contains NGC~5248 for which 
we clearly confirm the known UCNR of radius 95\,pc 
(Laine et al. 2001; Maoz et al. 2001; Jogee et al. 2002). 
We thus find UCNRs in four
of our 38 galaxies, which, if extrapolated, implies that 10\% of
spiral galaxies may contain one or more UCNRs---this value of course 
should be taken with caution as the present sample is neither complete nor
unbiased. The sample is biased in favour of disk galaxies
which show a central drop in the stellar velocity dispersion---hence
they may have a dynamically young stellar population inside the central kpc.
 
The new UCNRs presented here resemble closely their larger counterparts
(e.g., Knapen 2005). Their
morphology: round, with knots of massive SF spread around the
ring, and with structured dust lanes reaching from the
outside, but with little dust structure inside the ring, fits the
generic picture of nuclear rings in which the gas, driven
inward under the influence of a bar (or other non-axisymmetric)
potential, slows down near one or more ILRs. The UCNRs look very much
like ``normal'' nuclear rings, albeit much smaller than usual.  

This may either mean that (1) the mass distribution in the host galaxy
is shallower, which places the ILRs further in, that (2)
there is a small-scale non-axisymmetry in the mass distribution that
tumbles with a fast pattern speed, or that
(3) there is another, presently unknown mechanism, which operates in
the vicinity of the SMBHs, e.g., of a purely hydrodynamical or
radiatively-hydrodynamical nature.  The first option implies that
UCNRs prefer galaxies with small bulges, which is apparently not borne 
out by the classifications of NGC~2985 and NGC~4800: Sab and Sb. The 
last possibility means that there is a physical difference in the
origin of at least some of the UCNRs and the ``normal'' nuclear rings.

In favour of a distinct UCNR population from the standard nuclear ring
population, speaks also the size distribution of the latter. Figure~8
in Laine et al. (2002) displays a sharp decrease in the number of 
nuclear rings with sizes smaller than $\sim 1$\,kpc. This decrease
is not caused by the resolution problem---even the smaller rings
in Laine et al. are above the resolution limit of typical 
ground-based imaging
of typical host galaxies.

\section{Concluding remarks}

In this paper, we present the discovery of UCNRs in three
spiral galaxies, identified from {\it HST} archival UV and \ha\
imaging, which, for the first time, include LINERs. Overall,
four galaxies among our sample of 38 display this phenomenon. 
Although the sample is neither complete nor unbiased, this rough
estimate puts the frequency of UCNRs at $\sim 10\%$. Moreover,
they are found now in Sy~2s, LINERs and non-AGN galaxies---hence their origin lies in galactic dynamical processes and not
in radiative processes related to the AGN activity. 
The UCNRs found range in radius from 30 to 130\,pc, and lie
in galaxies of types Sab and Sb.

Two of our galaxies host UCNRs that look very much like classical
nuclear rings and are probably resonance rings, formed as a
consequence of gas inflow slowing down near one or more ILRs. One,
NGC~2985, has close and influencing companions, while the other,
NGC~4800, is weakly barred.  The UCNR in the third galaxy, NGC~4579,
is off-centre which can be explained as a result of a
superposition of $m=1$ and $m=2$ perturbations. The inner arcs
detected in this object appear to have a natural explanation in the
combined emission from the jet-induced SF and the
postshock gas.

The small number of UCNRs discovered does not yet allow us to reach
any conclusions on their formation or evolution as a class. However,
these results suggest  that there is no qualitative difference
between these rings and the nuclear rings seen in disk galaxies,
although they may constitute a separate population from the kpc-size
rings.
More UCNRs are expected to be found in our ongoing search, which
will allow us to establish the fraction of galaxies with UCNRs, and to
study in detail the relations to their host galaxies. Given the
proximity of UCNRs to central SMBHs, these findings will be of great
interest to the issue of fuelling of AGN activity, as well as to
that of galactic dynamics on scales of tens to a hundred pc.

\acknowledgements We thank Leonel Guti\'errez for his help with the
image processing and helpful ideas, and Tom Oosterloo and Edo Noordermeer for discussions on the \hi{} in NGC~2985. We also thank the referee, Herv\'e Wozniak. Support by the Ministerio de
Educaci\'on y Ciencia (AYA 2004-08251-CO2-01), the Instituto de
Astrof\'\i sica de Canarias (P3/86 and 3I2407), NSF and NASA is
gratefully acknowledged. Based on observations made with the NASA/ESA
{\it HST}, obtained from the data archive at the STScI, which is
operated by AURA under NASA contract NAS 5-26555.

\label{lastpage}

\end{document}